\documentclass[aps,pre,preprint,superscriptaddress,showpacs,amssymb]{revtex4}
\usepackage{graphicx}
\usepackage{graphics}
\usepackage{amsmath}
\usepackage{tabularx}
\usepackage{color}
\usepackage{doi}
\usepackage{textcomp}

\begin{document}

\title{Epidemic Outbreaks on Random Delaunay Triangulations}

\author{T. F. A. Alves}
\affiliation{Departamento de F\'{\i}sica, Universidade Federal do Piau\'{i}, 57072-970, Teresina - PI, Brazil}
\author{G. A. Alves}
\affiliation{Departamento de F\'{i}sica, Universidade Estadual do Piau\'{i}, 64002-150, Teresina - PI, Brazil}
\author{A. Macedo-Filho}
\affiliation{Campus Prof.\ Antonio Geovanne Alves de Sousa, Universidade Estadual do Piau\'i, 64260-000, Piripiri - PI, Brazil}
\author{R. S. Ferreira}
\affiliation{Departamento de Ci\^{e}ncias Exatas e Aplicadas, Universidade Federal de Ouro Preto, 35931-008, Ouro Preto - MG, Brazil}

\date{Received: date / Revised version: date}

\begin{abstract}

We study epidemic outbreaks on random Delaunay triangulations by applying Asynchronous SIR (susceptible-infected-removed) model kinetic Monte Carlo dynamics coupled to lattices extracted from the triangulations. In order to investigate the critical behavior of the model, we obtain the cluster size distribution by using Newman-Ziff algorithm, allowing to simulate random inhomogeneous lattices and measure any desired percolation observable. We numerically calculate the order parameter, defined as the wrapping cluster density, the mean cluster size, and Binder cumulant ratio defined for percolation in order to estimate the epidemic threshold. Our findings suggest that the system falls into two-dimensional dynamic percolation universality class and the quenched random disorder is irrelevant, in agreement with results for classical percolation.

\end{abstract}


\maketitle

\section{Introduction}

In this present work, we consider the Asynchronous SIR (susceptible-infected-removed) model\cite{Bernoulli-1760, Othsuki-1986, Renshaw-1991, Keeling-2007, Henkel2008, Tome-2010, deSouza-2011, Tome-2011, Tome-2015, Ruziska-2017, Pastorsatorras2015} coupled to random two-dimensional (2D) Delaunay triangulations. Random Delaunay triangulations are a simple way to introduce spatial disorder in any stochastic model, conceived to simulate a spreading disease in a population\cite{Renshaw-1991, Dickman1999, Keeling-2007, Tome-2015} while being embedded in space. By considering random Delaunay triangulations in a stochastic model by extracting its respective lattices, one implements spatial disorder in the form of connectivity disorder, defined as the feature of random lattices whose coordination numbers are not constant from node to node.

Random spatial disorder arising from defects or impurities can be found in most real systems, which we can refer as disordered systems. However, we should differentiate between two types of spatial disorder, depending on typical time scales (see, for example, \cite{Janke2002} and references therein). If the time scale of any dynamical process in the ``pure'' system (defined as the higher symmetrical system without the spatial disorder) is clearly separated from the time scale of spatial rearrangements, we can take as a good approximation the defects or impurities as frozen along the dynamical process. In this case, we have the quenched disorder. For the converse, where the dynamical processes and spatial rearrangements of the system are on the same time scale, i.e., the random degrees of freedom being ergodic, we have the annealing disorder. Here, we considered spatial disorder as a quenched disorder, which means we should simulate a number of system replicas, and then, relevant averages are done on both dynamics and quench realizations.

The understanding of disorder effects on the critical behavior of physical systems is a central topic in the statistical physics of phase transitions\cite{Janke2002}, including non equilibrium ones\cite{Dickman1999, Henkel2008, Tome-2015}. An important question is if disorder can affect universality, and some criteria have been formulated trying to answer this question. For relevant random disorder a system can change its critical behavior according to Harris\cite{ABHarris1974} criterion which states that quenched random disorder is relevant if
\begin{equation}
d\nu_\perp < 2,
\label{harriscriterion}
\end{equation}
where $d$ is the system dimensionality and $\nu_\perp$ is the correlation length exponent of the ``pure'' model. However, many physical systems have defects which are spatially correlated, in a way a uncorrelated and isotropic defect distribution is not adequate to describe these systems. Examples of these systems are the quasicrystals which possess quasiperiodic order\cite{Shechtman1984}. The generalization of Harris criterion\cite{ABHarris1974} for quasiperiodic long-range order is Luck criterion\cite{Luck1993}, which states that quasiperiodic long-range order is relevant if the wandering exponent $\omega$ that measures geometrical fluctuations of the lattice, satisfies $\omega > \omega_c$, where the marginal value for the wandering exponent $\omega_c$ is given by
\begin{equation}
\omega_c = 1 - \frac{1}{d\nu_\perp}.
\label{harris-luck-criterion}
\end{equation}
One should note that Harris criterion\cite{ABHarris1974} is recovered for uncorrelated random disorder with a wandering exponent $\omega = 0.5$\cite{Janke2004}. However, it is known that Harris-Luck criterion is violated, for example, the CP model coupled to 2D random Delaunay lattices\cite{Oliveira2008}, and Ising model on three-dimensional random Delaunay lattices\cite{Janke2002}. To accommodate these puzzling results in a more general scheme, Harris-Luck criterion was superseded by Harris-Barghathi-Vojta criterion\cite{Vojta2009,Barghathi2014}, which states that the connectivity disorder is relevant if
\begin{equation}
a\nu < 1,
\label{harris-barghathi-vojta-criterion}
\end{equation}
where $a$ is the lattice disorder decay exponent, defined by the scaling relation
\begin{equation}
\sigma = L^{-a}.
\end{equation}
Here, $\sigma$ is the coordination number deviation from the asymptotic mean coordination number. According to Harris-Barghathi-Vojta criterion, quasiperiodic tilings have irrelevant disorder in both Delaunay triangulation lattices and quasiperiodic tilings. However, even the modified criterion is known to be violated\cite{Schrauth-2018}.

In the lack of a general criterion to predict if connectivity disorder can change the critical behavior of a particular model, our objective is to investigate if random uncorrelated disorder is relevant for the Asynchronous SIR model coupled to random Delaunay triangulations. To accomplish that, we obtained the relevant parameters: the average percolating cluster density, the average mean cluster size, and Binder cumulant defined for percolation by using the Newman-Ziff algorithm\cite{Newman-2001}, and estimated the epidemic threshold and three independent critical exponent ratios: $1/\nu$, $\beta/\nu$ and $\gamma/\nu$. It is known that the Asynchronous SIR model can display a continuous phase transition between an endemic state to an epidemic state in the dynamic percolation universality class\cite{Othsuki-1986, Pastorsatorras2015, Tome-2010, deSouza-2011} for 2D periodic and quasiperiodic lattices\cite{Alves-SIR-quasiperiodic}.

The paper is outlined as follows. In section II, we describe kinetic Monte Carlo rules of Asynchronous SIR model dynamics, the method we used to build 2D random Delaunay triangulations, and the calculated parameters by Newman-Ziff algorithm from clusters grown by dynamics. In section III, we show and discuss our numerical results for the average infinite cluster density, the average mean cluster size, Binder cumulant, and our numerical estimations for the critical exponent ratios $1/\nu$, $\beta/\nu$ and $\gamma/\nu$. Finally, in section IV, we present our conclusions.

\section{Model and Implementation}

SIR model can be expressed as a reaction-diffusion process\cite{vanKampen-1981} whose lattice dynamics has the following rules:
\begin{itemize}
\item [(1)] We begin with a population of $N$ individuals, each one attached to its respective lattice node. Every node can assume one of three states: susceptible (S), infected (I), or removed (R). Dynamics begins with only one randomly chosen infected node known as the ``patient zero'' and all the other nodes at the susceptible state. We update a list of infected individuals and a list of removed individuals along the dynamics. The infected list starts with the ``patient zero'' and the list of removed sites starts empty;
\item [(2)] Next, we randomly choose an infected node $i$ from the infected list and proceed as follows:
	\begin{itemize}
	\item[(a)] We generate a random uniform number $x$ in the interval $[0,1)$. If $x \leq \lambda$, then we let the infected node become a removed node. We update the infected list and removed list accordingly;
	\item[(b)] If $x > \lambda$, we randomly choose a neighboring node and if the selected node is a susceptible one, we let it become infected and update the infected list accordingly;
	\end{itemize}
\item[(3)] We repeat step (2) several times until the system reaches any absorbing state, where there is not any infected nodes. For each step, we can increment a clock by a time lapse $1/N_i$, where $N_i$ is the number of infected particles, in an asynchronous fashion. The control parameter $\lambda$ is identified as the recovery rate and the infection rate equals $1-\lambda$.
\end{itemize}

SIR dynamics always evolve to an absorbing state, where there are not infected nodes. However, clusters formed by the activated nodes, i.e., nodes that became infected and eventually removed, can wrap the lattice, defining an epidemic phase. When we do not have a wrapping cluster, we have the endemic phase. A wrapping cluster is analogous of a percolating cluster in the infinite lattice for periodic boundaries in a way that the epidemic phase is analogous of a percolating phase for classical percolation. Following classical percolation theory\cite{Stauffer-1992, Christensen-2005}, we need the cluster distribution $n_{\mathrm{cluster}}(s)$, i.e., the number of clusters with $s$ removed vertexes in order to calculate percolation observables. The cluster distribution can be numerically obtained by using Newman-Ziff algorithm\cite{Newman-2001} with the feature of identifying if there is a wrapping cluster resulting from the dynamics.

By exploiting the fact that the epidemic phase is associated with the existence of a percolating cluster\cite{deSouza-2011,Alves-SIR-quasiperiodic}, we identify the order parameter as the average wrapping cluster density, i.e., the average fraction of activated nodes belonging to the wrapping cluster
\begin{equation}
P = \left[ \left< P_\infty \right> \right].
\label{orderparameter}
\end{equation}
Here, brackets $\left< \cdots \right>$ mean an average done over dynamics realizations and brackets $\left[ \cdots \right]$ indicate a quench average, done on random lattice realizations. From the finite cluster distribution, we can obtain another relevant parameter which is the mean cluster size $S$
\begin{equation}
S = \frac{1}{N_r}\sum_s s^2 n_{\mathrm{cluster}}(s),
\label{mean-cluster-size}
\end{equation}
whose average is a quantity analogous to the susceptibility of spin models
\begin{equation}
\chi = \left[ \left< S \right> \right],
\label{susceptibility}
\end{equation}
which we will refer as average mean cluster size. Following reference \cite{deSouza-2011}, we can define a ratio analogous to Binder cumulants for spin models, given by
\begin{equation}
U = \left[ \left< P_\infty \right> \frac{ \left< M' \right>}{ \left< S' \right> ^2} \right],
\label{bindercumulant}
\end{equation}
where observables $S'$ and $M'$ are written as
\begin{eqnarray}
S' &=& \frac{1}{N_r}\left(s^2_\mathrm{perc} + \sum_s s^2 n_{\mathrm{cluster}}(s)\right), \\
M' &=& \frac{1}{N_r}\left(s^3_\mathrm{perc} + \sum_s s^3 n_{\mathrm{cluster}}(s)\right),
\label{s-m}
\end{eqnarray}
with $s_\mathrm{perc}$ being the size of the percolating cluster. In the infinite size lattice limit, $s_\mathrm{perc} \rightarrow \infty$ and in the endemic phase, $s_{\mathrm{perc}}=0$, in a way observables $S'$ and $M'$ should be evaluated only at finite lattices.

SIR model dynamics have a particular feature: each realization of dynamics grows only one cluster of size $s=N_r$\cite{deSouza-2011, Alves-SIR-quasiperiodic}, therefore we have $n_{\mathrm{cluster}}(s)=\delta_{s,N_r}$ ($\delta$ is a Kronecker index) and $s_\mathrm{perc}=0$ at the endemic phase. For the epidemic (percolating) phase we should have $n_{\mathrm{cluster}}(s)=0$ and $s_\mathrm{perc}=N_r$. Then, observables $S'$ and $M'$ reduce to $S'=N_r$ and $M'=N_r^2$, and the order parameter reduces to $P_\infty=0$ if there is not a percolating cluster and to $P_\infty=1$ for the converse.

The simple cluster structure resulting from SIR model dynamics can be simulated without Newman-Ziff algorithm, exploiting lattice homogeneity and non-periodic boundaries, growing the cluster from a defined lattice center for every dynamics realization. In this simulation scheme, a percolating cluster is defined as the cluster that reaches any node in the lattice border grown from lattice center\cite{deSouza-2011}. However, it is a question if this simple scheme can be applied on non-homogeneous lattices with non-constant coordination numbers. Newman-Ziff algorithm\cite{Newman-2001} can be used to make the simulation of SIR dynamics applicable to non-homogeneous lattices when we should randomly choose the ``patient zero'' because of lattice inhomogeneities by dealing with them with an ensemble. Another advantage is the possibility to implement periodic boundaries\cite{Yang-2012}, reducing finite size effects\cite{Newman-2001} as discussed in the following section.

Relevant observables in Eqs. (\ref{orderparameter}), (\ref{susceptibility}), and (\ref{bindercumulant}) should scale as\cite{Stauffer-1992, Christensen-2005, deSouza-2011}
\begin{eqnarray}
P &\approx& L^{-\beta/\nu}f_{P}\left( L^{1/\nu} \left| \lambda - \lambda_c \right| \right), \label{orderparameter-fss} \\
\chi &\approx& L^{\gamma/\nu}f_\chi\left( L^{1/\nu} \left| \lambda - \lambda_c \right| \right), \label{susceptibility-fss} \\
U &\approx& f_U\left( L^{1/\nu} \left| \lambda - \lambda_c \right| \right), \label{bindercumulant-fss}
\end{eqnarray}
and, in an analogous way of Binder cumulant for spin models, curve crossings of $U$ ratio for different lattice sizes should give an estimate for the epidemic threshold $\lambda_c$, and scaling relations on Eqs. (\ref{orderparameter-fss}), (\ref{susceptibility-fss}), and (\ref{bindercumulant-fss}) can be used to estimate the critical exponents $\beta$, $\gamma$ and $\nu$ by standard data collapses, respectively, to completely determine the universality class.

\begin{figure}[h!]
\begin{center}
\includegraphics[scale=0.23]{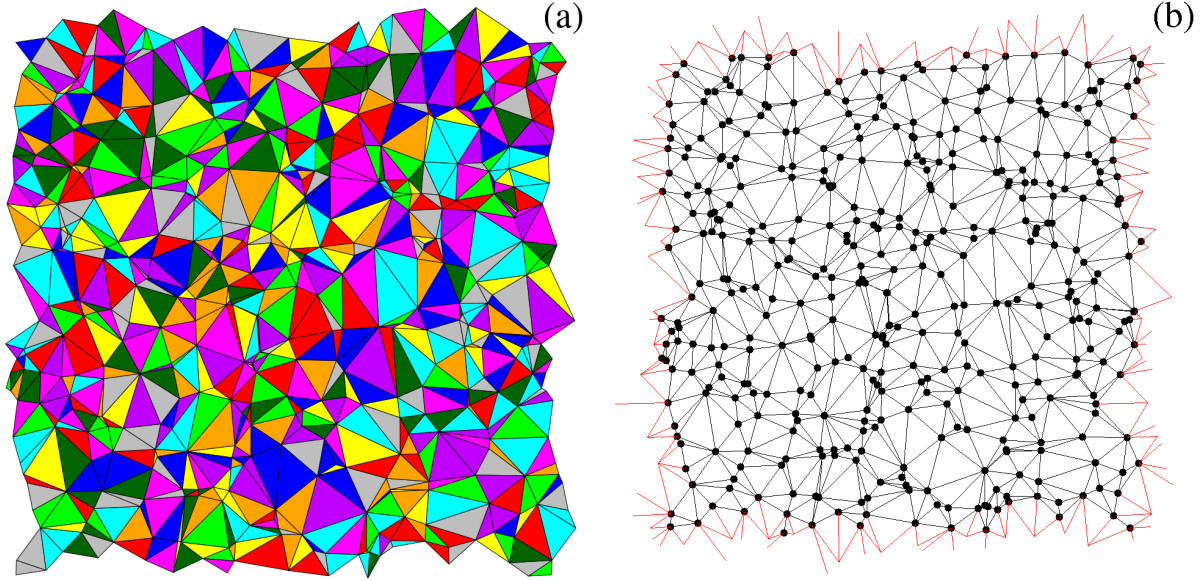}
\end{center}
\caption{(Color Online) (a) Delaunay triangulation generated by our implementation of Bowyer-Watson algorithm for a cloud of $N=400$ random points inside of a primary square domain with area $L^2=N$. Note the periodic boundaries, with matching triangle edges at opposed margins of the tiling. Periodic boundaries are implemented by creating full copies of the primary domain in the eight neighboring domains, totalizing $9N$ points to triangulate in a bigger square domain of area $9L^2$. (b) Delaunay random lattice extracted from the tiling (a). In black we have bulk edges and in grey (red), we have the periodic boundary edges connecting nodes across the boundaries. To identify boundary edges, one can search by edges connecting a vertex inside the primary domain and another vertex inside a neighbor domain.}
\label{delaunay-lattice}
\end{figure}

We coupled Asynchronous SIR model to 2D random Delaunay lattices, formed from random Delaunay triangulations. One example of a random Delaunay triangulation is shown in Fig.(\ref{delaunay-lattice}a) and its respective lattice is shown in Fig.(\ref{delaunay-lattice}b). 2D random Delaunay triangulations are a way of tiling the plane, where we have a cloud of $N$ random points, with $x$ and $y$ coordinates uniformly distributed in the interval $[0,L)$, inside a domain composed of a square of area $L^2$. For the sake of simplicity, we define the domain size as $L=\sqrt{N}$. Triangles are formed in such a way that no vertex point is inside the circumcircle of any triangle in the triangulation.

\begin{figure}[h!]
\begin{center}
\includegraphics[scale=0.25]{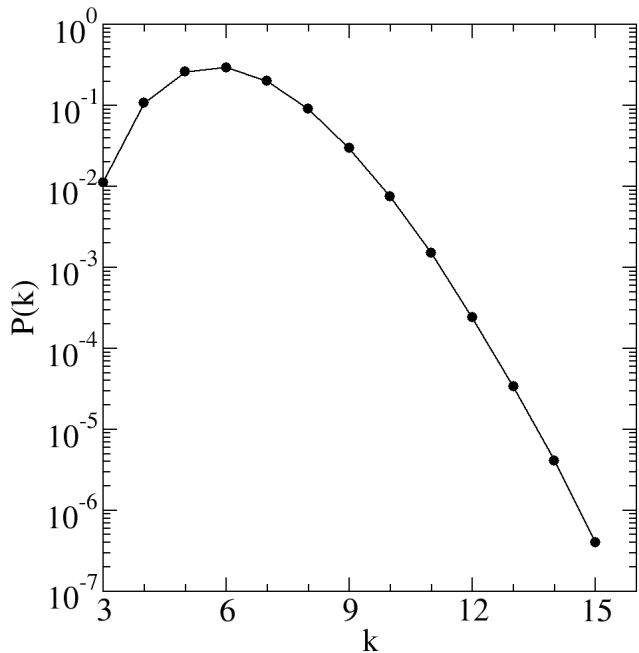}
\end{center}
\caption{Degree distribution $P(k)$ of 2D random Delaunay triangulations, obtained by generating $2000$ lattice realizations of size $L=100$ by using our implementation of Bowyer-Watson algorithm\cite{Bowyer-1981, Watson-1981}, where $k$ is the coordination number, i.e., number of neighbors of a particular node.}
\label{voronoi-degree}
\end{figure}

To generate the triangulations, we made use of Bowyer-Watson algorithm\cite{Bowyer-1981, Watson-1981} which is a method to generate Delaunay triangulations in any dimensions. It is an incremental algorithm that takes $O(N \ln N)$ operations to triangulate $N$ points. We slightly modified Bowyer-Watson algorithm to implement periodic boundaries by creating full copies of the cloud in the primary domain at the eight neighboring domains, totalizing $9N$ points to triangulate in a bigger square domain of area $9L^2$. Optionally we can only make copies of the points close to the boundaries by introducing a threshold on the coordinates to triangulate less points at the incremental process, exploiting the fact that the periodic boundary edges connect only points close to the boundaries as seen from Fig.(\ref{delaunay-lattice}b). To test if the lattice generated by our implementation of Bowyer-Watson algorithm have the desired properties, we obtained the degree distribution $P(k)$ showed in Fig.(\ref{voronoi-degree}), by generating $2000$ lattice realizations of domain size $L=100$, which reproduces Fig.(2) of reference \cite{Oliveira2008}.

In order to obtain numerical averages to make data collapses and to present the relevant averages as functions of the recovery rate, we repeated SIR model dynamics $10^{5}$ times for every lattice realization, starting from a randomly chosen ``patient zero'' until the system evolved to an absorbing state, growing $10^{5}$ clusters. We calculated an ensemble composed of $10^{5}$ measurements of order parameter, mean cluster size and Binder cumulant, to take the relevant ensemble averages for each lattice realization. Statistical errors for each lattice realization were calculated by using the ``jackknife'' resampling\cite{Tukey1958}. We repeated this procedure for $128$ random lattice realizations and finally made a quench average on the ensemble averages and statistical errors.

We followed the same route to generate data for the finite size scaling regressions, with the exception of the number of clusters grown for each lattice realization. For finite size scaling regressions, we repeated SIR dynamics $10^{6}$ times, generating an ensemble of $10^{6}$ observables to give better estimates of the observables at the epidemic threshold with smaller statistical errors. In the following section, we discuss our obtained numerical results.

\section{Results and Discussions}

\begin{figure}[p]
\begin{center}
\includegraphics[scale=0.16]{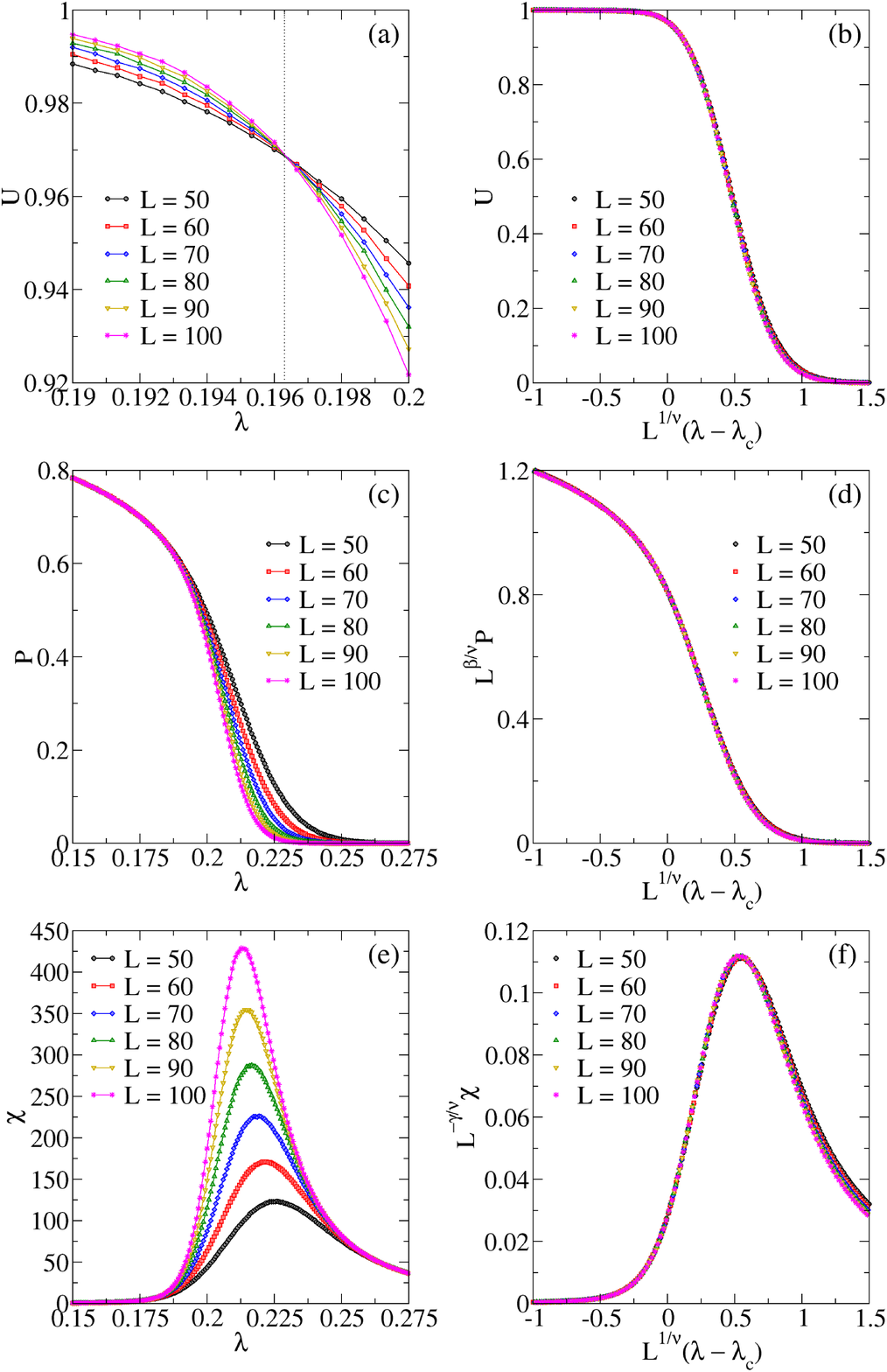}
\end{center}
\caption{In panels (a), (c), and (e), we show our numerical data for the Binder cumulant $U$, average wrapping cluster density $P$ and average mean cluster size $\chi$ as functions of recovery rate $\lambda$ for 2D random Delaunay lattices, respectively. In panels (b), (d), and (f), we show our numerical data for the Binder cumulant $U$, average wrapping cluster density $P$ and average mean cluster size $\chi$ rescaled by using the finite size scaling relations on Eqs. (\ref{orderparameter-fss}), (\ref{susceptibility-fss}), and (\ref{bindercumulant-fss}) with the 2D dynamic percolation exponent ratios $1/\nu=3/4$, $\beta/\nu=5/48$, $\gamma/\nu=43/24$, and the epidemic threshold $\lambda_c \approx 0.1963(3)$, respectively. Statistical errors are smaller than symbols.}
\label{observables-fig}
\end{figure}

In this section, we show our numerical results of Asynchronous SIR model coupled to 2D random Delaunay lattices. We begin our discussion by showing results of Binder cumulant in Fig.(\ref{observables-fig}a) and from the approximated crossing location, we estimate the epidemic threshold at $\lambda_c = 0.1963(3)$. Following Binder cumulant, we show results for order parameter in Fig.(\ref{observables-fig}c) whose typical sigmoidal shape indicates a continuous phase transition from the epidemic phase to the endemic phase. As seen in Fig.(\ref{observables-fig}e), the averaged mean cluster size diverges in the infinite size lattice limit, and its maxima gets closer to the epidemic threshold $\lambda_c$ when increasing $L$, as expected for a continuous phase transition.

We collapse all our numerical data shown in Figs.(\ref{observables-fig}a), (\ref{observables-fig}c) and (\ref{observables-fig}e) in Figs.(\ref{observables-fig}b), (\ref{observables-fig}d) and (\ref{observables-fig}f), by using the finite size scaling relations written in Eqs. (\ref{orderparameter-fss}), (\ref{susceptibility-fss}), and (\ref{bindercumulant-fss}), respectively, with the known exact values of the 2D dynamic percolation exponent ratios: $1/\nu=3/4$, $\beta/\nu=5/48$, and $\gamma/\nu=43/24$. Our data collapses suggesting that the system belongs to the 2D dynamic percolation universality class, irrespective of quenched random disorder present on 2D random Delaunay triangulations. This result is in agreement with classical percolation results for the same lattice\cite{Becker-2009}.

\begin{figure}[h!]
\begin{center}
\includegraphics[scale=0.15]{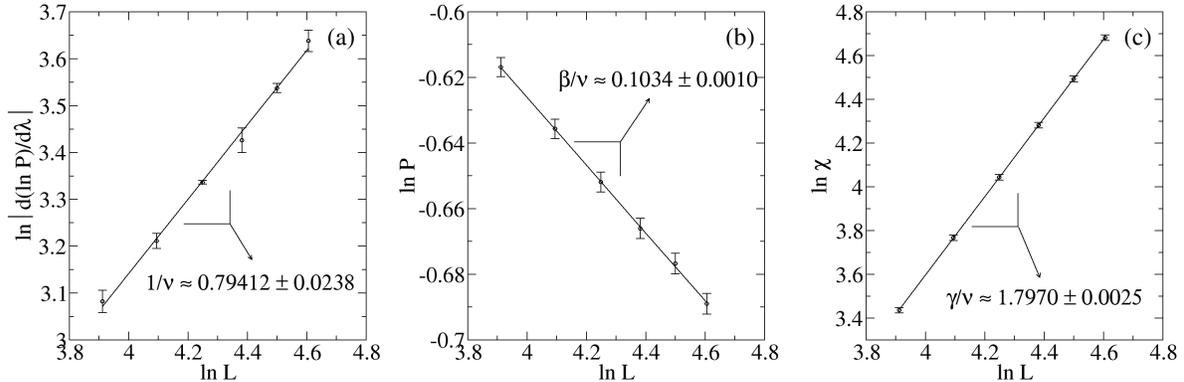}
\end{center}
\caption{In panels (a), (b), and (c), we show estimates by finite size scaling linear regressions of $\ln \left|\frac{\mathrm{d}\left(\ln{P}\right)}{\mathrm{d}\lambda}\right|$, $\ln P$, and $\ln \chi$, respectively, as functions of $\ln L$ for 2D random Delaunay lattices at the epidemic threshold $\lambda_c = 0.1963(3)$. All estimated exponent ratios are close to the respective 2D dynamic percolation exponents $1/\nu=3/4$, $\beta/\nu=5/48$, and $\gamma/\nu=43/24$. Note the logarithm scale for the data and statistical errors.}
\label{regressions-fig}
\end{figure}

Finally, we show estimates for the critical exponent ratios $1/\nu$, $\beta/\nu$, and $\gamma/\nu$ by using finite size scaling regressions of $\ln \left| \frac{\mathrm{d}\left(\ln{P}\right)}{\mathrm{d}\lambda}\right|$, $\ln P$, and $\ln \chi$ shown on Figs.(\ref{regressions-fig}a), (\ref{regressions-fig}b), and (\ref{regressions-fig}c), respectively, as functions of $\ln L$. All observables were evaluated at the critical threshold $\lambda_c \approx 0.1963$. Error bars of $\left| \frac{\mathrm{d}\left(\ln{P}\right)}{\mathrm{d}\lambda}\right|$ are extracted from the least squares method applied to data points close to the epidemic threshold, used to obtain the logarithm derivative data.

Our estimates for the critical exponent ratios are $1/\nu \approx 0.7941 \pm 0.0238$, $\beta/\nu \approx 0.1034 \pm 0.0010$, and $\gamma/\nu \approx 1.7970 \pm 0.0025$. Numerical value of $1/\nu$ from the finite size scaling regression deviates $4\%$ from the exact critical exponent ratio $1/\nu=3/4$, and numerical values of $\beta/\nu$ and $\gamma/\nu$ deviate less than $1\%$ from exact $\beta/\nu=5/48$ and $\gamma/\nu=43/24$ values of the 2D dynamical percolation universality class. We see that the periodic boundary conditions enables us to achieve a much better numerical precision in the numerical estimation of critical exponents in comparison with our results for the same model on quasiperiodic lattices\cite{Alves-SIR-quasiperiodic}. The same conclusion was made on reference \cite{Newman-2001} and the explanation is given in terms of smaller finite size perturbations.

\section{Conclusions}

We proposed the use of Newman-Ziff algorithm combined to the kinetic Monte Carlo dynamics of Asynchronous SIR model to numerically determine the cluster size distribution and the wrapping cluster for periodic boundaries in order to calculate any desired observable related to percolation. By following a new scheme of simulating this present model on non-homogeneous random lattices embedded in space like 2D random Delaunay lattices, we were able to identify the epidemic phase of an epidemic outbreak model. We also were able to obtain its order parameter defined as the average wrapping cluster density, the average mean cluster size and a cumulant ratio defined for percolation in order to estimate the epidemic threshold at $\lambda_c = 0.1963(3)$. We investigated the critical behavior of the model and our numerical results suggest that the random quenched disorder is irrelevant and the system falls into the 2D dynamical percolation universality class in the same way of the Asynchronous SIR model on periodic and quasiperiodic lattices, in agreement with classical percolation results.

\section{Acknowledgments}

We would like to thank CNPq (Conselho Nacional de Desenvolvimento Cient\'{\i}fico e tecnol\'{o}gico), FUNCAP (Funda\c{c}\~{a}o Cearense de Apoio ao Desenvolvimento Cient\'{\i}fico e Tecnol\'{o}gico) and FAPEPI (Funda\c{c}\~{a}o de Amparo a Pesquisa do Estado do Piau\'{\i}) for the financial support. We acknowledge the use of Dietrich Stauffer Computational Physics Lab, Teresina, Brazil, and Laborat\'{o}rio de F\'{\i}sica Te\'{o}rica e Modelagem Computacional - LFTMC, Piripiri, Brazil, where the numerical simulations were performed.

\bibliography{textv1}

\end{document}